\documentclass[twocolumn,amsmath,amssymb,eps]{revtex4}

\usepackage{graphicx,epsfig}
\usepackage{dcolumn}
\usepackage{bm}
\newcommand{\beq}{\begin{equation}}
\newcommand{\eeq}{\end{equation}}
\newcommand{\bseq}{\begin{subequations}}
\newcommand{\eseq}{\end{subequations}}
\newcommand{\beqa}{\begin{eqnarray}}
\newcommand{\eeqa}{\end{eqnarray}}
\newcommand{\nnbar}{n{\bar n}}

\begin{document}
\title{Multi-GeV neutrinos due to $\nnbar$ oscillation in Gamma-Ray Burst 
Fireballs}
\author{Sarira Sahu}

\affiliation{
Instituto de Ciencias Nucleares, Universidad Nacional Aut\'onoma de M\'exico, 
Circuito Exterior, C.U., A. Postal 70-543, 04510 Mexico DF, Mexico
}

\begin{abstract}

The long and short gamma-ray bursts are believed to be produced due to
collapse of massive stars and merger of compact binaries respectively. 
All these objects are rich in neutron and the jet outflow from these objects 
must have a neutron component in it. By postulating the  $\nnbar$ oscillation
in the gamma-ray burst fireball, we show that, 19-38 GeV neutrinos and 
anti-neutrinos can be produced due to annihilation of anti-neutrons with
the background neutrons. These neutrinos and anti-neutrinos will be 
produced before the 5-10 GeV neutrinos due to dynamical decoupling of
neutrons from the rest of the fireball. Observation of these neutrinos
will shed more light on the nature of the GRB progenitors and  also be a
unique signature of physics beyond the standard model. A
possible way of detecting these neutrinos in future is also discussed.

\end{abstract}

\maketitle

\section{Introduction}

Gamma Ray Bursts (GRBs) are flashes of non-thermal bursts of low energy 
($\sim$ 100 KeV-1 MeV) photons and release about $10^{51}$-$10^{53}$ erg in a 
few seconds making them the most luminous object in the universe. They are
known to occur at cosmological distance\cite{meszaros1,piran}. They fall into 
two classes; short-hard bursts ($\le 2~s$) and long-soft bursts. The short 
bursts comprises about 25\% of the total events and the rest are 
long-soft bursts. It is now widely accepted that long duration bursts are 
produced due to the core collapse of massive stars the so called hypernovae.
The origin of short-duration bursts are still a mystery, but recently there 
has been tremendous progress  due to accurate localization of many short
bursts by the Swift\cite{gehrels,barthelmy} and HETE-2\cite{villasenor} 
satellites. The afterglow observation of
GRB 050709 at a redshift of 0.1606\cite{hjorth} by HETE-2 and the Swift 
observation of afterglow from GRB050709b at a redshift of 0.225\cite{gehrels} 
and GRB 050724 at a redshift of 0.258\cite{berger} seems to support the 
coalescing of compact binaries as the progenitor for the short-hard bursts 
although definite conclusions can not be drawn at this stage.

Irrespective of the nature of the progenitor, it is believed that, gamma-ray
emission arrises from the collision of different internal shocks (shells) due 
to relativistic outflow from the source. A class 
of models call {\it fireball model} seems to explain the temporal structure
of the bursts and the non-thermal nature of their 
spectra\cite{piran,meszaros1}. 
Sudden release of copious amount of 
$\gamma$ rays into a compact region with a
size $c\delta t\sim 100$-$1000$ km\cite{piran} 
creates an opaque $\gamma-e^{\pm}$ fireball
due to the process $\gamma+\gamma\rightarrow e^+ + e^-$. The average optical
depth of this process\cite{guilbert} is 
$\tau_{\gamma\gamma}\simeq 10^{13}$.
This optical depth is very large and even if there are no pairs to 
begin with, they will form very 
rapidly and will Compton scatter lower energy photons. Because of the huge
optical depth, photons can not escape freely. In the fireball the $\gamma$ and
$e^{\pm}$ pairs will thermalize with a temperature of about 3-10 MeV.
The fireball  expands relativistically with a Lorentz factor 
$\Gamma\sim 100-1000$  under its
own pressure and cools adiabatically due to the expansion. The radiation 
emerges freely to the inter galactic medium (ISM), 
when the optical depth is $\tau_{\gamma\gamma}\simeq 1$. 

In addition to $\gamma$, $e^{\pm}$ pairs, fireballs may also contain some
baryons, both from the progenitor and the surrounding medium.
These baryons can be either free or in the form of nuclei. 
If the fireball temperature
is high enough (more than 0.7 MeV), then it will be mostly in the form of 
neutrons and protons. 
The electrons associated with the matter (baryons) can 
increase the opacity, hence delaying the process of emission of radiation and
the baryons can be accelerated along with the fireball and convert part of the
radiation energy into bulk kinetic energy.
So the dynamics of the fireball crucially depends on the baryon content of it.
But irrespective of it, the baryonic load has to be very 
small, otherwise, the expansion of the fireball will be
Newtonian, which is inconsistent with the present observations. 

As discussed above, core collapse of massive stars and merger of binary
compact objects (neutron star-neutron star, neutron star-black hole) as
possible progenitors of the long and short GRBs respectively. In both the
cases neutrons are the main ejected materials from the central engine. So it
seems obvious that, the outflow from the central engine, which later on give
rise to GRBs, has neutron as a principal component. The presence of neutron
component in GRBs was first proposed by Derishev et al,\cite{derishev1} 
and later on it was
showed that the presence of neutron component in the GRBs are practically
inevitable and its presence drastically changes the dynamics of the
fireball. The possible role of neutron in observed emission has been discussed
by many authors\cite{derishev1,derishev2,beloborodov,bahcall,meszaros2}. 
The dynamical decoupling of neutron from the rest of the relativistic
shell will give rise to inelastic n-p scattering and lead to emission
of observable multi-GeV neutrinos\cite{bahcall,meszaros2}.

It is believed that, baryon number is not a good symmetry of nature, due to
the fact that, our universe is matter dominant and this requires 
baryon number violating interactions. Thus the breaking of exact-baryon number
conservation in unified theories of the fundamental gauge
interactions\cite{pati} has lead
to searches for  proton decay and $\nnbar$ oscillation, but so far with no
success\cite{mohapatra}. The $\nnbar$ oscillation occurs due to 
$\Delta B=2$ transition. The present
experimental limit for unbound neutrons is 
$\tau_{\nnbar}\sim 10^9~s$\cite{kamyshkov}. 

The neutron oscillation in free space and its effect on cosmic rays has been 
discussed in Ref.\cite{sawada}.  As a
phenomenological implication of $\nnbar$ oscillation, 
it has been suggested that, the excess
of sub-GeV anti-protons over protons in the galactic cosmic rays is due to the
$\nnbar$ oscillation in supernova explosions, which are site for large neutron
excess\cite{sivaram1}. Also it has been argued that, $\nnbar$ oscillation put 
a lower limit on the primordial magnetic field when applied to the neutron rich
environment of the nucleosynthesis era\cite{sivaram2}.

Here we propose that, if $\nnbar$ oscillations take place within 
the neutron rich GRB fireball, $19-38~GeV$ neutrinos will be produced prior
to the  production of 5-10 GeV neutrinos due to dynamical neutron decoupling 
from the rest of the fireball as predicted by Bahcall and 
M\'esz\'aros\cite{bahcall}. By using earth as the detector and assuming the
magnetic field in the jet outflow from the central engine to be $\sim
10^{-6}~G$, about 4 events can be recorded per year. The future
international project Extreme Universe Space Observatory (EUSO) may be
able to detect these neutrinos\cite{euso}.

\section{$\nnbar$ Oscillation and Physics of GRBs}

In the rest frame, neutron has a mean life time of 
$\tau_{\beta}=888.6\pm3.5~s$. In the comoving frame, when the neutron has a 
Lorentz factor $\Gamma_n$, the mean decay radius is 
$r_{\beta}=\Gamma_n \tau_{\beta}$. So due to beta decay, at a distance $r$
from the source, the neutron number is given by
\beq
N_n=N_{n,0} e^{-r/r_{\beta}},
\eeq
where $N_{n,0}$ is original number of neutron within a radius 
$r\sim r_{\beta}\sim 10^6~cm$. In the presence of a magnetic field, the 
neutron energy levels are split by an amount $\Delta E = g\mu B$ and this 
is responsible for the oscillation of neutron to anti-neutron, where 
$g=-1.91$ is the anomalous magnetic moment of neutron, 
$\mu=e\hbar/2 m_p c=3.152\times 10^{-12} ~ 
eV~ G^{-1}$ and $B$ is the magnetic field in Gauss. Due to $\nnbar$ 
oscillation, the number of anti-neutron is given by\cite{sawada}
\beq
N_{\bar n}\simeq \frac{1}{2} {N_n}
\left (\frac{\delta m}{ \Delta E} \right )^2=0.6\times 10^{-26}N_n~\left (
\frac{B}{G}\right )^{-2}.
\label{neutronbar}
\eeq
where $\delta m={\tau_{\nnbar}}^{-1}$ and here we take 
$\tau_{\nnbar}\simeq 10^{9} s$. The number of ${\bar n}$ is inversely
proportional to the square of the magnetic field. So for large magnetic field
this is very much suppressed and for small one it is enhanced.

In the fireball, a substantial fraction of the baryon kinetic energy is
transfered to a non thermal population of electrons through Fermi acceleration
at the shock and these accelerated electrons cool via synchrotron emission
and/or inverse Compton scattering to produce the observed prompt (due to
internal shocks) and afterglow (due to external shocks) emission. For the
synchrotron emission, strong magnetic field is needed to fit the observed
data. But it is difficult to estimate the strength of the magnetic field from
the first principle. One would expect large magnetic field if the progenitors
are highly magnetized, for example, magnetars with $B\sim 10^{16}~G$. Also a
relatively small pre-existing magnetic field can be amplified due to turbulent
dynamo mechanism, compression or shearing. Despite all these, there is no 
satisfactory explanation for the
existence of strong magnetic field in the fireball. Also even if some magnetic
flux is carried by the outflow, it will decrease due to the expansion of the
fireball at a larger radius. Recently, it has been proposed that, 
the emission in the GRBs can also be explained through Compton-drag process 
and magnetic field is not necessary for the production of gamma rays
\cite{liang,lazzati,ghisellini}. Because of the above uncertainties, here we
consider the magnetic field  as a parameter.

By combining both the neutron decay and the $\nnbar$ oscillation,
the number of anti-neutrons
at a distance $r$ from the source are given by
\beqa
{N_{\bar n}} &\simeq & \frac{1}{2}\left ( \frac{\xi}{1+\xi}\right )
\frac{E}{\eta m_p} e^{-r/r_{\beta}} 
\left (\frac{\delta m}{ \Delta E} \right )^2\nonumber\\
&=&  2\times 10^{27}
\left (\frac{B}{G}\right )^{-2}\left ( \frac{2\xi}{1+\xi}\right )
\frac{E_{53}}{\eta_{100}} e^{-r/r_{\beta}}, 
\label{antin}
\eeqa
where $\xi$ is the neutron to proton ratio at the source and for simplicity
we take $\xi\sim 1$ and the dimensionless entropy  $\eta=E/M\sim 10^2-10^3$, 
where $E$ is the initial radiation energy produced due to $e^{\pm},\gamma$ and
$M$ is the rest mass of the fireball due to baryon load. 

The protons, neutrons and electrons are coupled to the thermal radiation in
the expanding  jet outflow from the central engine until the Compton 
scattering
time scale $t^{\prime}_{Th}\simeq (n^{\prime}_p\sigma_{Th})^{-1}$ and the
elastic n-p scattering time scale $t^{\prime}_{np}
\simeq (n^{\prime}_{p}\sigma_{np})^{-1}$ are shorter  than the comoving
plasma expansion time scale $t^{\prime}_{exp}\simeq  r/\Gamma$, where 
$\sigma_{np}\sim 3\times 10^{-26}~cm^2$ is the n-p elastic 
scattering cross section, $n^{\prime}_{p}$ is the comoving number density of
protons in the fireball. In fact when all the components in the jet are
coupled, they have a common bulk Lorentz factor 
$\Gamma\sim 300$ Ref.\cite{bahcall}.
The neutrons and protons are coupled until the opacity 
$\tau_{np} >1$, which corresponds to the $r_{np}$ radius
\beq
r_{np} <\frac{L\sigma_{np}}{(1+\xi) 4\pi m_p\Gamma^2\eta}
\sim 6\times 10^{10} \frac{L_{52}}{(1+\xi) \Gamma^2_{300} \eta_{100}}~ 
cm.
\eeq
So the radius $r$ for the n-p to be coupled should be smaller than $r_{np}$, 
i.e. $r < r_{np}$. Also the mean beta decay radius is $r_{\beta} 
\sim  10^{16}~ cm >> r$. So $ {exp}(-r/r_{\beta})\sim 1$  
and we can comfortably neglect  the effect of beta decay in 
Eq.(\ref{antin}) when we are in the 
regime $ r<r_{np}<r_{\beta}$.
 
In the jet frame protons and neutrons have the same Lorentz factor
$\Gamma$ and the $\nnbar$ oscillation takes place with in the
equilibrium plasma. The produced anti-neutrons can annihilate with the 
background neutrons to produce pions and also can have elastic
scattering (but not inelastic scattering) with the background
protons. The beta decay of anti-neutrons will be suppressed due to the
longer life time of the anti-neutrons in the comoving jet plasma
unless the decay takes place in the regime $ r<r_{np}<r_{\beta}$. We
consider the following annihilation processes within the comoving plasma,
\beqa
{\bar n} + n&\rightarrow& \pi^+ +\pi^- +\pi^0\nonumber\\
{\bar n} + n&\rightarrow& \pi^+ +\pi^- \nonumber\\
{\bar n} + n&\rightarrow& \pi^0+\pi^0.
\label{nbarn}
\eeqa 
The $\pi^0$ will decay to $2\gamma$ and the charge pions will decay as, 
\beq
\pi^{\pm}\rightarrow \mu^{\pm}+\nu_{\mu} ({\bar \nu}_{\mu})
\rightarrow e^{\pm}+\nu_e ({\bar \nu}_e) +\nu_{\mu} ({\bar \nu}_{\mu})+
{\bar\nu}_{\mu} (\nu_{\mu}).
\label{pipmnu}
\eeq
In Eq.(\ref{nbarn}) the energy carried by each pion in the c.m. frame is 
$1.88 ~GeV/k_{\pi}$ , where $k_{\pi}$ is the multiplicity of pion. 
The produced pions are
relativistic and in the first step of Eq.(\ref{pipmnu}), the charged pions 
will decay to
$\mu^{\pm}+ \nu_{\mu}({\bar\nu_{\mu}})$.  In these decays, the average energy
carried by muon is  $\sim 80\%$ of the pion energy and rest is
carried by $\nu_{\mu}({\bar\nu_{\mu}})$\cite{gaisser}. 
In the second step of the decay
muon will decay as shown in Eq.(\ref{pipmnu}) and the average energy carried by
each particle is about 1/3 of the muon energy. So on an average the energy
carried by each neutrino or anti-neutrino due to muon decay is 
$\epsilon^{\prime}_{\nu,{\bar\nu}}\simeq 0.5 GeV/k_{\pi}$ in the 
comoving frame. 
In the observer frame this will be translated to
\beq
\epsilon_{\nu,{\bar\nu}}\simeq 0.5~ \frac{GeV}{k_{\pi}} ~\Gamma~
\frac{1}{(1+z)}.
\eeq
By normalizing the burst at redshift $z=1$, we obtain
\beq
\epsilon_{\nu,{\bar\nu}}\simeq 75~ 
\frac{GeV}{k_{\pi}} ~ \Gamma_{300}~ \frac{2}{1+z}.
\eeq
For the pion multiplicity $k_{\pi}=3$ and $2$, this gives 
$\epsilon_{\nu,{\bar\nu}}=25~GeV$ and $37.5~GeV$ respectively. Similarly
the energy of the muon neutrinos due to pion decay
$\pi^{\pm}\rightarrow\mu^{\pm}+ \nu_{\mu}({\bar\nu_{\mu}})$ is
\beq 
\epsilon_{\nu_{\mu},{\bar\nu}_{\mu}}\simeq 56~GeV/k_{\pi}~ 
\Gamma_{300}~\frac{2}{1+z}, 
\eeq
which is $19~ GeV$ and $28~GeV$ respectively for
$k_{\pi}=3$ and $2$. So the neutrinos and anti-neutrinos produced
due to $n\bar n$ annihilation are having energies in the range $19~GeV$ to
$38~GeV$. 

The neutral pions will decay to photons and have energies $0.3~GeV$ (first of 
Eq.(\ref{nbarn})) and $0.47~GeV$ (last of Eq.(\ref{nbarn})) respectively in
the c.m. frame. In the observer frame, these will be boosted to $45~GeV$ and
$71~GeV$ respectively. As we have discussed earlier, 
these photons are produced at a radius $r<r_{np}$ and $\tau_{\gamma\gamma} >
1$, they will degrade their energy by producing $e^{\pm}$ pairs and thermalize
within the fireball medium. But the neutrinos and anti-neutrinos 
which are produced due
to $n{\bar n}$ annihilation will stream away from the fireball much before 
the decoupling of neutrons. So these neutrinos are different from the
one that are produced due to neutron decoupling in the later stage of
the fireball evolution\cite{bahcall}. 
The total amount of energy released due to $n\bar n$ annihilation is 
\beq
\epsilon_{n\bar n}=2m_n N_{\bar n}\simeq 6\times 10^{24}~E_{53}\eta^{-1}_{100}
\left ( \frac{B}{G}\right )^{-2} 
\left (\frac{2\xi}{1+\xi}\right )~erg,
\eeq
and a major fraction of it will be emitted in the form of neutrinos and 
anti-neutrinos. 
 
\section{Possible number of neutrino events}

By considering both the long and short GRBs rate within a Hubble radius of 
$R_b\sim 10^5R_{b5}$\cite{waxman}, the number of events per year in a 
detector with $N_t$ target protons  is
\beq 
R_{\nu {\bar\nu}}\sim ({N_t/4\pi D^2}) R_b N_{\bar n}
{{\bar\sigma}_{\nu{\bar\nu}}},
\eeq 
where 
\beq
D=2.64\times 10^{28} h^{-1}_{70} (1-1/\sqrt{1+z})~cm,
\eeq
is the proper distance 
out to red shift $z$ with $h_{70}=H_0/70 ~km/s/Mpc$ and
${{\bar\sigma}_{\nu{\bar\nu}}}\sim 5\times 10^{-40} \Gamma (1+z)^{-1}~cm^2$ 
is the total detection cross section per neutron. 
If we consider the whole earth as the detector with $N_t\sim 10^{51}$ and 
taking a small magnetic field $B\sim 10^{-6}~G$ in the jet outflow (in
Eq.(\ref{antin})), the event rate is 
\beqa
R_{\nu {\bar\nu}} & \sim & 
4~ h^2_{70} R_{b5} \frac{E_{53}}{\eta_{100}} N_{t51}\Gamma_{300}   
\left (\frac{B}{10^{-6}G}\right )^{-2}\nonumber\\
&&\times \left ( \frac{2\xi}{1+\xi}\right )
\left(\frac{3-2\sqrt{2}}{2+z-2\sqrt{1+z}}
\right)year^{-1}.
\eeqa
So the expected rate of multi-GeV neutrino event is about 4 per year and
the increase in the initial neutron to proton ratio $\xi$
will slightly increase the event rate\cite{raz}.
May be in future, the EUSO will be able to detect these neutrinos
by emission of fluorescence light of nitrogen due to the extensive air showers 
through the interaction of the multi-GeV neutrinos with the
atmosphere\cite{euso}. 
  
\section{Conclusions}

The production of neutrinos and anti-neutrinos due to $n{\bar n}$ annihilation
is inversely proportional to the square of the magnetic field.  If the
magnetic field in the fireball is very weak, the number of anti-neutrons
will be more as can be seen from  Eq.(\ref{neutronbar}). The produced
anti-neutrons will annhilate with the
background neutrons through the processes in Eq.(\ref{nbarn}) to produce
neutrinos, anti-neutrons and photons. A large fraction of the
annihilation energy will stream away in the form of neutrinos, thus
reducing the energy content of the fireball. 
For example if the magnetic field in the fireball is very small ($B\simeq
10^{-12}~G$), the amount of energy release due to $n\bar n$ annihilation is 
$\epsilon_{n\bar n}\sim 10^{49}~ erg$. 
Apart from the $\nnbar$ annihilation, if beta decay of
anti-neutrons take place within the region  $ r < r_{np}$ with a
very small magnetic field, the excess amount of baryon,
anti-baryon annihilation will result in decreasing the baryon content
of the fireball further and making it cleaner, which can probably explain the
almost baryon free environment of the fireball. 
On the other hand, strong magnetic field will suppress the $\nnbar$ 
oscillation and ultimately produce very less number of neutrinos and 
anti-neutrinos. So if the $\nnbar$ oscillation exist in GRB fireballs, 
magnetic field can not be arbitrarily weak or strong.
The produced neutrinos and anti-neutrinos are unique signature of
$\nnbar$ oscillation  as well as the presence of magnetic field 
in the GRB fireballs.  It is unique in the sense that, there is no other
process which can produce these neutrinos in between the production point
(the source at $r=r_0$ ) and $r=r_{np}$. Flux of these neutrinos
depends on the original neutron contain as well as the magnetic
field in the fireball. Due to very low flux, these neutrinos will be very 
difficult to detect. The future project EUSO may be able to
detect these neutrinos. However, if at all
detected, it will be before the 5-10 GeV neutrinos due to inelastic
scattering of decoupling neutrons with the protons in the fireball. 
The observed time difference between the $\nnbar$ neutrinos and the 
5-10 GeV neutrinos is  
$\Delta t\simeq r_{np}/\Gamma^2\sim 4\times 10^{-6}~s$. Although this
time difference is very small, the energies of the neutrinos are very
different. So the neutrinos within the energy range 19-38 GeV have 
different physics (production mechanism) as compared to 5-10 GeV
neutrinos. Observation of these multi-GeV neutrinos will not only tell more
about the nature of the GRB progenitors but also shed more light on the
physics beyond the standard model. 

The author is thankful to J. F. Nieves for helpful discussion.
This research is partially supported by DGAPA-UNAM project IN119405.



\begin{thebibliography}{100}
\bibitem{meszaros1} P. M\'esz\'aros, astro-ph/0605208; B. Zhang and
  P. M\'esz\'aros, {\it IJMPA} {\bf 19}, 2385 (2004). 
\bibitem{piran} T. Piran, {\it Phys. Rep.} {\bf 314}, 575 (1999).
\bibitem{gehrels} N. Gehrels, et al., {\it Nature} {\bf 437}, 851 (2005).
\bibitem{barthelmy} S. D. Barthelmy, et al., {\it Nature 438}, 994 (2005).
\bibitem{villasenor} J. S. Villasenor, et al., {\it Nature} {\bf 437}, 855
  (2005). 
\bibitem{hjorth} J. Hjorth, et al., {\it Nature} {\bf 437}, 859 (2005).
\bibitem{berger} E. Berger, et al., {\it Nature} {\bf 438}, 988 (2005).
\bibitem{guilbert} P. W. Guilbert, A. C. Fabian and M. J. Rees, {\it
  Mon. Not. R. Astr. Soc.} {\bf 205}, 593 (1983).
\bibitem{derishev1} E. V. Derishev, V. V. Kocharovsky and VL. V.  Kocharovsky,
{\it Astro. Phys. J.} {\bf 521}, 640 (1999). 
\bibitem{derishev2} E. V. Derishev, V. V. Kocharovsky and VI. V.  Kocharovsky,
{\it Astron. Astrophys.} {\bf 345}, L51 (1999).
\bibitem{beloborodov} A. M. Beloborodov, {\it Astro. Phys. J.}, {\bf 588}, 931
  (2003). 
\bibitem{bahcall} J. N. Bahcall and P. M\'esz\'aros, {\it Phys. Rev. Lett.} 
{\bf 85}, 1362 (2000). 
\bibitem{meszaros2} P. M\'esz\'aros and M. J. Rees, {\it Astro. Phys. J.} {\bf
 541}, L5 (2000). 

\bibitem{pati} J. C. Pati and A. Salam, {\it Phys. Rev. Lett.} {\bf 31}, 661
  (1973); H. Georgi and S. L. Glashow, {\it Phys. Rev. Lett.} {\bf 32}, 438
  (1974). 
\bibitem{mohapatra} S. L. Glashow, {\it Proc. Neutrino 79}, 
Bergen, Vol. {\bf 1},
  p. 518 (1979); R. N. Mohapatra and R. E. Marshak, {\it Phys. Rev. Lett.}
  {\bf 44}, 1316 (1980); V. A. Ku\'zmin, {\it JETP Lett.} {\bf 12}, 228
  (1970); R. N. Mohapatra, hep-ph/0605289.
\bibitem{kamyshkov} Yu. Kamyshkov, {\it Nucl. Phys. B} (Proc. Suppl.) 
{\bf 52A}, 263 (1997).
\bibitem{sawada} O. Sawada and M. Fukugita, {\it Astro. Phys. J.} {\bf 248}, 
1162 (1981).
\bibitem{sivaram1} C. Sivaram and V. Krishan, {\it Nature} {\bf 299}, 
427 (1982). 
\bibitem{sivaram2}  C. Sivaram and V. Krishan, {\it Astrophys. Lett} {\bf 23},
139 (1983).
\bibitem{euso} See EUSO Web page, http://www.euso-mission.org/.
\bibitem{liang} E. P. Liang, {\it Astro. Phys. J.} {\bf 491}, L15 (1997). 
\bibitem{lazzati} D. Lazzati and G. Ghisellini, {\it Astro. Phys. J.} 
{\bf 529}, L17 (2000).
\bibitem{ghisellini} G. Ghisellini and A. Celotti, {\it Astro. Phys. J.} {\bf
  511}, L93 (1999).
\bibitem{gaisser} T. K. Gaisser, Cosmic Rays and Particle Physics, Cambridge 
University Press, Cambridge (1990).
\bibitem{waxman} P. M\'esz\'aros and E. Waxman,{\it Phys. Rev. Lett.} {\bf
  87}, 171101 (2001). 
\bibitem{raz} S. Razzaque and P. M\'esz\'aros, {\it Astro. Phys. J.} {\bf
  650}, 998 (2006). 
\end{thebibliography}
\end{document}